\documentclass[journal]{IEEEtran}

\usepackage{amsfonts}
\usepackage{amsmath}
\usepackage{amssymb}
\usepackage{lscape}
\usepackage{epsf}
\usepackage{graphicx}
\usepackage{psfrag}

\newtheorem{theorem}{Theorem}
\newtheorem{lemma}{Lemma}
{\indent Claim}
\newtheorem{EXAMPLE}{Example}
\newtheorem{definition}{Definition}

\newcommand{\cM}{{\mathcal{M}}}

\newcommand{\coeff}{{\mbox{Coeff }}}

\newcommand{\bldalpha}{{\mbox{\boldmath $\alpha$}}}
\newcommand{\bldsmallalpha}{{\mbox{\scriptsize \boldmath $\alpha$}}}

\newcommand{\bldbeta}{{\mbox{\boldmath $\beta$}}}
\newcommand{\bldsmallbeta}{{\mbox{\scriptsize \boldmath $\beta$}}}

\newcommand{\bldepsilon}{{\mbox{\boldmath $\epsilon$}}}
\newcommand{\bldsmallepsilon}{{\mbox{\scriptsize \boldmath $\epsilon$}}}

    \def\squarebox#1{\hbox to #1{\hfill\vbox to #1{\vfill}}}

\newlength{\Algwidth}

\title{Growth Rate of the Weight Distribution of Doubly-Generalized LDPC Codes: General Case and Efficient Evaluation
\thanks{%
    M. F. Flanagan is with the School of Electrical, Electronic and Mechanical Engineering, University College Dublin, Belfield, Dublin 4, Ireland (e-mail:mark.flanagan@ieee.org).
    \newline
    E. Paolini and M. Chiani are with DEIS, University of Bologna, Via Venezia 52, 47023 Cesena (FC), Cesena, Italy (e-mail:e.paolini@unibo.it, marco.chiani@unibo.it).
    \newline
    M. P. C. Fossorier is with ETIS ENSEA, UCP, CNRS UMR-8051, 6 avenue du Ponceau, 95014 Cergy Pontoise, France (e-mail: mfossorier@ieee.org). 
    }        
}

\author{Mark F. Flanagan, Enrico Paolini, Marco Chiani, and~Marc~P.~C.~Fossorier}

\begin{document}
\maketitle
\thispagestyle{empty}

\begin{abstract}
The growth rate of the weight distribution of irregular doubly-generalized LDPC (D-GLDPC) codes is developed and in the process, a new efficient numerical technique for its evaluation is presented. The solution involves simultaneous solution of a $4 \times 4$ system of polynomial equations. This represents the first efficient numerical technique for \emph{exact} evaluation of the growth rate, even for LDPC codes. The technique is applied to two example D-GLDPC code ensembles. 
\end{abstract}

\section{Introduction}
Recently, the design and analysis of coding schemes representing generalizations of Gallager's low-density parity-check (LDPC) codes \cite{gallager63:low-density} has gained increasing attention. This interest is motivated above all by the potential capability of these coding schemes to offer a better compromise between waterfall and error floor performance than is currently offered by state-of-the-art LDPC codes.

In the Tanner graph of an LDPC code, any degree-$q$ variable node (VN) may be interpreted as a length-$q$ repetition code, i.e., as a $(q,1)$ linear block code. Similarly, any degree-$s$ check node (CN) may be interpreted as a length-$s$ single parity-check (SPC) code, i.e., as a $(s,s-1)$ linear block code. The first proposal of a class of linear block codes generalizing LDPC codes may be found in \cite{tanner81:recursive}, where it was suggested to replace each CN of a regular LDPC code with a generic linear block code, to enhance the overall minimum distance. The corresponding coding scheme is known as a regular generalized LDPC (GLDPC) code, or Tanner code, and a CN that is not a SPC code as a generalized CN. More recently, irregular GLDPC codes were considered (see for instance \cite{liva08:quasi_cyclic}). For such codes, the VNs exhibit different degrees and the CN set is composed of a mixture of different linear block codes.

A further generalization step is represented by doubly-generalized LDPC (D-GLDPC) codes \cite{wang06:D-GLDPC}. In a D-GLDPC code, not only the CNs but also the VNs may be represented by generic linear block codes. The VNs which are not repetition codes are called generalized VNs. The main motivation for introducing generalized VNs is to overcome some problems connected with the use of generalized CNs, such as an overall code rate loss which makes GLDPC codes interesting mainly for low code rate applications, and a loss in terms of decoding threshold (for a discussion on drawbacks of generalized CNs and on beneficial effects of generalized VNs we refer to \cite{miladinovic08:generalized} and \cite{paolini09:stability}, respectively).

A useful tool for analysis and design of LDPC codes and their generalizations is represented by the growth rate of the weight distribution, or equivalently, the asymptotic weight enumerating function (WEF). The growth rate of the weight distribution was introduced in \cite{gallager63:low-density} to show that the minimum distance of a randomly generated regular LDPC code with a VN degree of at least three is a linear function of the codeword length with high probability. The same approach was taken in \cite{lentmaier99:generalized} and \cite{boutros99:generalized} to obtain related results on the minimum distance of subclasses of Tanner codes.

The growth rate of the weight distribution has been subsequently investigated for unstructured ensembles of irregular LDPC codes. Works in this area are \cite{litsyn02:on_ensembles,burshtein04:asymptotic,orlitsky05:stopping,di06:weight}. In particular, in \cite{di06:weight} a technique for approximate evaluation of the growth rate of any (eventually expurgated) irregular LDPC ensemble has been developed, based on Hayman's formula. Asymptotic weight enumerators of ensembles of irregular LDPC codes based on protographs and on multiple edge types have been derived in \cite{divsalar06:weight_enumerators} and \cite{awano08:three-edge_type}, respectively. The approach proposed in \cite{divsalar06:weight_enumerators} has then been extended to protograph GLDPC codes and to protograph D-GLDPC codes in \cite{abu-surra07:ensemble_GLDPC} and \cite{wang08:ensemble_DGLDPC}, respectively. In \cite{flanagan08:growth}, the authors presented a compact formula for the growth rate of general unstructured irregular D-GLDPC code ensembles for the specific case of \emph{small} weight codewords. 

In this paper, an analytical expression for the growth rate of the weight distribution of a general unstructured irregular ensemble of D-GLDPC codes is developed. As opposed to the formula developed in \cite{flanagan08:growth}, the proposed expression holds for any codeword weight. The present work also extends to the fully-irregular case an expression for the growth rate obtained in \cite{paolini09:class} assuming a CN set composed of linear block codes all of the same type. In the process of this development, we obtain an efficient evaluation tool for computing the growth rate exactly. This tool always requires the solution of a $(4\times4)$ polynomial system of equations, \emph{regardless} of the number of VN types and CN types in the D-GLDPC ensemble. As shown through numerical examples, the proposed tool allows to obtain a precise plot of the growth rate with a low computational effort. For the case of irregular LDPC codes, a  technique for numerical evaluation of the growth rate of the weight distribution was given in \cite{di06:weight}; in contrast to the technique developed in this paper, the method of \cite{di06:weight} provided an \emph{approximate} numerical solution for the growth rate; it is also more computationally complex than that proposed in the present work.


\section{Preliminaries and Notation}\label{section:irregular_D_GLDPC}
We define a D-GLDPC code ensemble $\cM_n$ as follows, where $n$ denotes the number of VNs. There are $n_c$ different CN types $t \in I_c = \{ 1,2,\cdots, n_c\}$, and $n_v$ different VN types $t \in I_v = \{ 1,2,\cdots, n_v\}$. For each CN type $t \in I_c$, we denote by $h_t$, $s_t$ and $r_t$ the CN dimension, length and minimum distance, respectively. For each VN type $t \in I_v$, we denote by $k_t$, $q_t$ and $p_t$ the VN dimension, length and minimum distance, respectively. For $t \in I_c$, $\rho_t$ denotes the fraction of edges connected to CNs of type $t$. Similarly, for $t \in I_v$, $\lambda_t$ denotes the fraction of edges connected to VNs of type $t$. Note that all of these variables are independent of $n$.

The polynomials $\rho(x)$ and $\lambda(x)$ are defined by $\rho(x) \triangleq \sum_{t\in I_c} \rho_t x^{s_t - 1}$ 
and $\lambda(x) \triangleq \sum_{t \in I_v} \lambda_t x^{q_t - 1}$. If $E$ denotes the number of edges in the Tanner graph, the number of CNs of type $t\in I_c$ is then given by $E \rho_t / s_t$, and the number of VNs of type $t\in I_v$ is then given by $E \lambda_t / q_t$. Denoting as usual $\int_0^1 \rho(x) \, {\rm d} x$ and $\int_0^1 \lambda(x) \, {\rm d} x$ by $\int \rho$ and $\int \lambda$ respectively, we see that the number of edges in the Tanner graph is given by $E =n / \int \lambda$ and the number of CNs is given by $m = E \int \rho$. Therefore, the fraction of CNs of type $t \in I_c$ and the fraction of VNs of type $t \in I_v$ are given by
\begin{equation}\label{eq:gamma_t_delta_t_definition}
\gamma_t = \frac{\rho_t}{s_t \int \rho} \quad \textrm{and} \quad \delta_t = \frac{\lambda_t}{q_t \int \lambda}
\end{equation}
respectively. Also the length of any D-GLDPC codeword in the ensemble is given by 
\begin{equation}
N = \sum_{t \in I_v} \left( \frac{E \lambda_t}{q_t} \right) k_t = \frac{n}{\int \lambda} \sum_{t \in I_v} \frac{\lambda_t k_t}{q_t} \; .
\label{eq:DG_LDPC_codeword_length}
\end{equation}
Note that this is a linear function of $n$. Similarly, the total number of parity-check equations for any D-GLDPC code in the ensemble is given by $M = \frac{m}{\int \rho} \sum_{t \in I_c} \frac{\rho_t (s_t - h_t)}{s_t}$. A member of the ensemble $\cM_n$ then corresponds to a permutation on the $E$ edges connecting CNs to VNs.

The WEF for CN type $t \in I_c$ is given by $A^{(t)}(z) = 1 + \sum_{u=r_t}^{s_t} A_u^{(t)} z^u$. 
Here $A_u^{(t)} \ge 0$ denotes the number of weight-$u$ codewords for CNs of type $t$. The input-output weight enumerating function (IO-WEF) for VN type $t \in I_v$ is given by $B^{(t)}(x,y) = 1 + \sum_{u=1}^{k_t} \sum_{v=p_t}^{q_t} B_{u,v}^{(t)} x^u y^v$.  
Here $B_{u,v}^{(t)} \ge 0$ denotes the number of weight-$v$ codewords generated by input words of weight $u$, for VNs of type $t$. Also, $B^{(t)}_{2}$ is the total number of weight-$2$ codewords for VNs of type $t$.

If there exist CNs and VNs with minimum distance equal to $2$, and define the (positive) parameters
\begin{equation}
C = 2 \sum_{t \; : \; r_t = 2} \frac{\rho_t A^{(t)}_{2}}{s_t} \: ; \: V = 2 \sum_{t \; : \; p_t = 2} \frac{\lambda_t B^{(t)}_{2}}{q_t} \; .
\label{eq:C_V_definitions}
\end{equation}

The design rate of any D-GLDPC ensemble is given by
\begin{equation}\label{eq:design_rate}
R = 1 - \frac{\sum_{t \in I_c} \rho_t (1 - R_t)}{\sum_{t \in I_v} \lambda_t R_t}
\end{equation}
where for $t \in I_c$ (resp. $t \in I_v$) $R_t$ is the local code rate of a type-$t$ CN (resp. VN). 

Throughout this paper, the notation $e = \exp(1)$ denotes Napier's number, all the logarithms are assumed to have base $e$ and for $0<x<1$ the notation $h(x)=-x \log(x) - (1-x) \log(1-x)$ denotes the binary entropy function. 


\section{Growth Rate of the Weight Distribution of General Irregular D-GLDPC Code Ensembles}\label{sec:growth_rate}
The growth rate of the weight distribution of the irregular D-GLDPC ensemble sequence $\{ \cM_n \}$ is defined by 
\begin{equation}
G(\alpha) \triangleq \lim_{n\rightarrow \infty} \frac{1}{n} \log \mathbb{E}_{\cM_n} \left[ N_{\alpha n} \right]
\label{eq:growth_rate_result}
\end{equation}
where $\mathbb{E}_{\cM_n}$ denotes the expectation operator over the ensemble $\cM_n$, and $N_{w}$ denotes the number of codewords of weight $w$ of a randomly chosen D-GLDPC code in the ensemble $\cM_n$. The limit in (\ref{eq:growth_rate_result}) assumes the inclusion of only those positive integers $n$ for which $\alpha n \in \mathbb{Z}$ and $\mathbb{E}_{\cM_n} [ N_{\alpha n} ]$ is positive. Note that the argument of the growth rate function $G(\alpha)$ is equal to the ratio of D-GLDPC codeword length to the number of VNs; by (\ref{eq:DG_LDPC_codeword_length}), this captures the behaviour of codewords linear in the block length, as in \cite{di06:weight} for the LDPC case. 

A D-GLDPC ensemble is said to be \emph{asymptotically good} if and only if $\alpha^* \triangleq \inf \{ \alpha > 0 \; | \; G(\alpha)\geq 0 \} > 0$. The parameter $\alpha^*$ is called the \emph{ensemble relative minimum distance}. In \cite{flanagan09:IEEE-IT}, it was shown that a D-GLDPC ensemble is always asymptotically good if there exist no CNs or VNs with minimum distance $2$ while, if the exist both CNs and VNs with minimum distance $2$, the ensemble is asymptotically good if and only if $C \cdot V < 1$, where $C$ and $V$ are given by (\ref{eq:C_V_definitions}).

Note that using (\ref{eq:DG_LDPC_codeword_length}), we may also define the growth rate with respect to the number of D-GLDPC code bits $N$ as follows:
\begin{equation}
H(\gamma) \triangleq \lim_{N\rightarrow \infty} \frac{1}{N} \log \mathbb{E}_{\cM_n} \left[ N_{\gamma N} \right] \; .
\label{eq:growth_rate_result_norm}
\end{equation}
It is straightforward to show that 
\begin{equation}
H(\gamma) = \frac{G(\gamma y)}{y}
\label{eq:relationship_G_H}
\end{equation}
where 
\[
y = \frac{1}{\int \lambda} \sum_{t \in I_v} \frac{\lambda_t k_t}{q_t} \; .
\]

In this section, we formulate an expression of the growth rate for an irregular D-GLDPC ensemble $\cM_n$ over a wider range of $\alpha$ than was considered in \cite{flanagan08:growth, flanagan09:IEEE-IT} (where the case $\alpha \rightarrow 0$ was analyzed).

The following theorem constitutes our main result.
\medskip
\begin{theorem}
The growth rate of the weight distribution of the irregular D-GLDPC ensemble sequence $\{ \cM_n \}$ is given by
\begin{multline}
G(\alpha) = \sum_{t \in I_v} \delta_t \log B^{(t)}(x_{0},y_{0}) - \alpha \log x_{0} \\
+ \left( \frac{\int \rho}{\int \lambda} \right) \sum_{s \in I_c} \gamma_s \log A^{(s)}(z_0) + \frac{\log \left( 1 - \beta \int \lambda \right)}{\int \lambda} 
\label{eq:growth_rate_polynomial_general}
\end{multline}
where $x_0$, $y_0$, $z_0$ and $\beta$ are the unique positive real solutions to the $4 \times 4$ system of polynomial equations\footnote{Note that while~(\ref{eq:z0_eqn}),~(\ref{eq:x0_y0_eqn_1}) and~(\ref{eq:x0_y0_eqn_2}) are not polynomial as set down here, each may be made polynomial by multiplying across by an appropriate factor.}
\begin{equation}
\label{eq:z0_eqn}
z_0 \left( \frac{\int \rho}{\int \lambda} \right) \sum_{t \in I_c} \gamma_t \frac{ \frac{\mathrm{d} A^{(t)}}{\mathrm{d} z} (z_{0})}{A^{(t)}(z_{0})} = \beta \; ,
\end{equation}
\begin{equation}
\label{eq:x0_y0_eqn_1}
x_0 \sum_{t \in I_v} \delta_t \frac{ \frac{\partial B^{(t)}}{\partial x} (x_{0},y_{0})}{B^{(t)}(x_{0},y_{0})} = \alpha \; ,
\end{equation}
\begin{equation}
\label{eq:x0_y0_eqn_2}
y_0 \sum_{t \in I_v} \delta_t \frac{ \frac{\partial B^{(t)}}{\partial y} (x_{0},y_{0})}{B^{(t)}(x_{0},y_{0})} = \beta \; ,
\end{equation}
and
\begin{equation}
\label{eq:z0_y0_relation}
\left( \beta \int \lambda \right) (1 + y_0 z_0) = y_0 z_0 \; .
\end{equation}
\label{thm:growth_rate} 
\end{theorem}
The theorem is proved in Section \ref{sec:proof_of_main_result}. 

\section{Proof of the Main Result}\label{sec:proof_of_main_result}
In this section we prove Theorem \ref{thm:growth_rate}. The proof uses the concepts of \emph{assignment} and \emph{split assignment}, defined next. 
\medskip
\begin{definition}
An \emph{assignment} is a subset of the edges of the Tanner graph. An assignment is said to have \emph{weight} $k$ if it has $k$ elements. An assignment is said to be \emph{check-valid} if the following condition holds: supposing that each edge of the assignment carries a $1$ and each of the other edges carries a $0$, each CN recognizes a valid local codeword.
\end{definition}
\medskip
\begin{definition}
A \emph{split assignment} is an assignment, together with a subset of the D-GLDPC code bits (called a \emph{codeword assignment}). A split assignment is said to have \emph{split weight} $(u, v)$ if its assignment has weight $v$ and its codeword assignment has $u$ elements. A split assignment is said to be \emph{check-valid} if its assignment is check-valid. A split assignment is said to be \emph{variable-valid} if the following condition holds: supposing that each edge of its assignment carries a $1$ and each of the other edges carries a $0$, and supposing that each D-GLDPC code bit in the codeword assigment is set to $1$ and each of the other code bits is set to $0$, each VN recognizes a local input word and the corresponding valid local codeword.     
\end{definition}
\medskip
For ease of presentation, the proof is broken into two parts.

\subsection{Number of check-valid assignments of weight $\delta m$}\label{subsec:no_of_check_valid_assignments}
First we derive an expression, valid asymptotically, for the number of check-valid assignments of weight $\delta m$. For each $t \in I_c$, let $\epsilon_t m$ denote the portion of the total weight $\delta m$ apportioned to CNs of type $t$. Then $\epsilon_t \ge 0$ for each $t \in I_c$, and $\sum_{t \in I_c} \epsilon_t = \delta$. Also denote $\bldepsilon = (\epsilon_1 \; \epsilon_2 \; \cdots \; \epsilon_{n_c})$. 

Consider the set of $\gamma_t m$ CNs of a particular type $t \in I_c$, where $\gamma_t$ is given by (\ref{eq:gamma_t_delta_t_definition}). Using generating functions, the number of check-valid assignments (over these CNs) of weight $\epsilon_t m$ is given by
\[
N_{c,t}^{(\gamma_t m)}(\epsilon_t m) = \coeff \left[ \left( A^{(t)}(x) \right) ^{\gamma_t m}, x^{\epsilon_t m} \right]
\]
where $\coeff [ p(x), x^c ]$ denotes the coefficient of $x^c$ in the polynomial $p(x)$. We next make use of the following result, which is a special case of \cite[Corollary 16]{di06:weight}:
\medskip
\begin{lemma}
Let $A(x) = 1 + \sum_{u=c}^{d} A_u x^u$, where $1 \le c \le d$, be a polynomial satisfying $A_c > 0$ and $A_u \ge 0$ for all $c < u \le d$. Then
\begin{equation}
\lim_{\ell\rightarrow \infty} \frac{1}{\ell} \log \coeff \left[ \left( A(x) \right) ^{\ell}, x^{\xi \ell} \right] = \log \left( \frac{A(z)}{z^{\xi}} \right)
\end{equation}
where $z$ is the unique positive real solution to
\begin{equation}
\label{eq:z_soln_to_A_eqn}
\frac{A'(z)}{A(z)} \cdot z = \xi \; .
\end{equation}
\label{lemma:optimization_1D}
\end{lemma} 
Applying this lemma by substituting $A(x) = A^{(t)}(x)$, $\ell=\gamma_t m$ and $\xi = \epsilon_t/\gamma_t$, we obtain that as $m \rightarrow \infty$
\begin{eqnarray}
N_{c,t}^{(\gamma_t m)}(\epsilon_t m) = \coeff \left[ \left( A^{(t)}(x) \right) ^{\gamma_t m}, x^{\epsilon_t m} \right] 
\label{eq:Nct_epsilon_start} \\
\rightarrow \exp \left\{ m \left( \gamma_t \log A^{(t)}(z_{0,t}) - \epsilon_t \log z_{0,t} \right) \right\} 
\label{eq:Nct_epsilon_mid}
\end{eqnarray}
where, for each $t \in I_c$, $z_{0,t}$ is the unique positive real solution to
\begin{equation}
\label{eq:z0t_soln_to_At_eqn}
\gamma_t \frac{ \frac{\mathrm{d} A^{(t)}}{\mathrm{d} z} (z_{0,t})}{A^{(t)}(z_{0,t})} \cdot z_{0,t} = \epsilon_t \; .
\end{equation}

The number of check-valid assignments of weight $\delta m$ satisfying the constraint $\bldepsilon$ is obtained by multiplying the numbers of check-valid assignments of weight $\epsilon_t m$ over $\gamma_t m$ CNs of type $t$, for each $t \in I_c$,
\begin{equation}
N_c^{(\bldepsilon)}(\delta m) = \prod_{t \in I_c} N_{c,t}^{(\gamma_t m)}(\epsilon_t m) \; .
\label{eq:Nc_epsilon}
\end{equation}
The number $N_c(\delta m)$ of check-valid assignments of weight $\delta m$ is then equal to the sum of $N_c^{(\bldepsilon)}(\delta m)$ over all admissible vectors $\bldepsilon$; therefore by~(\ref{eq:Nct_epsilon_mid}), as $m\rightarrow \infty$
\begin{equation}
N_c(\delta m) \rightarrow \sum_{\bldsmallepsilon \; : \; \sum_{t \in I_c} \epsilon_t = \delta}
\exp \left\{ m W(\bldepsilon) \right\}
\label{eq:sum_of_exp_check}
\end{equation}
where
\begin{equation}
W(\bldepsilon) = \sum_{t \in I_c} \left( \gamma_t \log A^{(t)}(z_{0,t}) - \epsilon_t \log z_{0,t} \right) \; .
\label{eq:W_definition}
\end{equation}
As $m \rightarrow \infty$, the asymptotic expression is dominated by the distribution $\bldepsilon$ which maximizes the argument of the exponential function\footnote{Observe that as $m\rightarrow \infty$, $\sum_t \exp ( m Z_t ) \rightarrow \exp ( m \max_t \{Z_t\} )$}. Therefore as $m\rightarrow \infty$
\begin{equation}
N_c(\delta m) \rightarrow \exp \left\{ m X \right\}
\label{eq:N_c_as_function_of_X}
\end{equation}
where 
\begin{equation}
X = \max_{\bldsmallepsilon} W(\bldepsilon)
\end{equation} 
and the maximization is subject to the constraint
\begin{equation}
V(\bldepsilon) = \sum_{t \in I_c} \epsilon_t = \delta 
\label{eq:sum_epsilont_constraint}
\end{equation}
together with $\epsilon_t \ge 0$ for each $t \in I_c$, and for every $t \in I_c$, $z_{0,t}$ is the unique positive real solution to~(\ref{eq:z0t_soln_to_At_eqn}). Note that for each $t \in I_c$, (\ref{eq:z0t_soln_to_At_eqn}) provides an implicit definition of $z_{0,t}$ as a function of $\epsilon_t$. 

We solve this optimization problem using Lagrange multipliers, ignoring for the moment the inequality constraints. At the maximum, we must have
\begin{equation}
\frac{\partial W(\bldepsilon)}{\partial \epsilon_t} = \lambda \frac{\partial V(\bldepsilon)}{\partial \epsilon_t}
\label{eq:max_condition_check}
\end{equation}
for all $t \in I_c$, where $\lambda$ is the Lagrange multiplier. This yields
\begin{equation}
\frac{\partial z_{0,t}}{\partial \epsilon_t} \left[ \gamma_t \frac{ \frac{\mathrm{d} A^{(t)}}{\mathrm{d} z} (z_{0,t})}{A^{(t)}(z_{0,t})} - \frac{\epsilon_t}{z_{0,t}} \right] - \log z_{0,t} = \lambda \; .
\end{equation}
The term in square brackets is equal to zero due to (\ref{eq:z0t_soln_to_At_eqn}); therefore this simplifies to $\log z_{0,t} = -\lambda$ for all $t \in I_c$.
We conclude that all of the $\{ z_{0,t} \}$ are equal, and we may write 
\begin{equation}
z_{0,t} = z_0 \quad \forall t \in I_c \; . 
\label{eq:z0_all_equal}
\end{equation}
Making this substitution in~(\ref{eq:N_c_as_function_of_X}) and using~(\ref{eq:sum_epsilont_constraint}) we obtain
\begin{equation}
N_c(\delta m) \rightarrow \exp \left\{ m \left( \sum_{t \in I_c} \gamma_t \log A^{(t)}(z_{0}) - \delta \log z_{0} \right) \right\} \; .
\label{eq:N_c_asymptotic}
\end{equation}
Summing~(\ref{eq:z0t_soln_to_At_eqn}) over $t \in I_c$ and using (\ref{eq:sum_epsilont_constraint}) and (\ref{eq:z0_all_equal}) implies that the value of $z_0$ in~(\ref{eq:N_c_asymptotic}) is the unique positive real solution to~(\ref{eq:z0_eqn}) (here we have also used the fact that $n \int \rho = m \int \lambda$).

\subsection{Polynomial-System Solution for the Growth Rate}\label{subsec:growth_rate}
Consider the set of $\delta_t n$ VNs of a particular type $t \in I_v$, where $\delta_t$ is given by (\ref{eq:gamma_t_delta_t_definition}). Using generating functions, the number of variable-valid split assignments (over these VNs) of split weight $(\alpha_t n, \beta_t n)$ is given by
\[
N_{v,t}^{(\delta_t n)}(\alpha_t n, \beta_t n) = \coeff  \left[ \left( B^{(t)}(x,y) \right) ^{\delta_t n}, x^{\alpha_t n} y^{\beta_t n} \right]
\]
where $\coeff [p(x,y), x^c y^d ]$ denotes the coefficient of $x^c y^d$ in the bivariate polynomial $p(x,y)$. We make use of the following result, which is a special case of \cite[Corollary 16]{di06:weight}:
\medskip
\begin{lemma}
Let 
\[
B(x,y) = 1 + \sum_{u=1}^{k} \sum_{v=c}^{d} B_{u,v} x^u y^v 
\]
where $k \ge 1$ and $1 \le c \le d$, be a bivariate polynomial satisfying $B_{u,v} \ge 0$ for all $1 \le u \le k$, $c \le v \le d$. Then
\begin{equation}
\lim_{\ell\rightarrow \infty} \frac{1}{\ell} \log \coeff \left[ \left( B(x,y) \right) ^{\ell}, x^{\xi \ell} y^{\theta \ell} \right] = \log \left( \frac{B(x_0,y_0)}{x_0^{\xi} y_0^{\theta}} \right)
\end{equation}
where $x_{0}$ and $y_{0}$ are the unique positive real solutions to the pair of simultaneous equations
\begin{equation}
\frac{ \frac{\partial B}{\partial x} (x_{0},y_{0})}{B(x_{0},y_{0})} \cdot x_{0} = \xi
\end{equation}
and
\begin{equation}
\frac{ \frac{\partial B}{\partial y} (x_{0},y_{0})}{B(x_{0},y_{0})} \cdot y_{0} = \theta \; . 
\end{equation}
\label{lemma:optimization_2D}
\end{lemma}
Applying this lemma by substituting $B(x,y) = B^{(t)}(x,y)$, $\ell = \delta_t n$, $\xi = \alpha_t/\delta_t$ and $\theta = \beta_t/\delta_t$, we obtain that as $n \rightarrow \infty$
\begin{eqnarray}
N_{v,t}^{(\delta_t n)}(\alpha_t n, \beta_t n) = \coeff  \left[ \left( B^{(t)}(x,y) \right) ^{\delta_t n}, x^{\alpha_t n} y^{\beta_t n} \right] \nonumber \\
\rightarrow \exp \left\{ n X_t^{(\delta_t)}(\alpha_t, \beta_t) \right\}
\label{eq:Nvt_asymptotic}
\end{eqnarray}
where 
\begin{equation}
X_t^{(\delta_t)}(\alpha_t, \beta_t) = \delta_t \log B^{(t)}(x_{0,t},y_{0,t}) - \alpha_t \log x_{0,t} - \beta_t \log y_{0,t} 
\end{equation}
and where $x_{0,t}$ and $y_{0,t}$ are the unique positive real solutions to the pair of simultaneous equations
\begin{equation}
\label{eq:x0t_y0t_eqn_1}
\delta_t \frac{ \frac{\partial B^{(t)}}{\partial x} (x_{0,t},y_{0,t})}{B^{(t)}(x_{0,t},y_{0,t})} \cdot x_{0,t} = \alpha_t
\end{equation}
and
\begin{equation}
\label{eq:x0t_y0t_eqn_2}
\delta_t \frac{ \frac{\partial B^{(t)}}{\partial y} (x_{0,t},y_{0,t})}{B^{(t)}(x_{0,t},y_{0,t})} \cdot y_{0,t} = \beta_t \; . 
\end{equation}

Next, note that the expected number of D-GLDPC codewords of weight $\alpha n$ in the ensemble $\cM_n$ is equal to the sum over $\beta$ of the expected numbers of split assignments of split weight $(\alpha n, \beta n)$ which are both check-valid and variable-valid, denoted $N^{v,c}_{\alpha n, \beta n}$:
\[
\mathbb{E}_{\cM_n} \left[ N_{\alpha n} \right] = \sum_{\beta} \mathbb{E}_{\cM_n} [ N^{v,c}_{\alpha n, \beta n} ] \; .
\]
This may then be expressed as
\begin{multline}
\mathbb{E}_{\cM_n} \left[ N_{\alpha n} \right] = \sum_{\substack{\alpha_t \ge 0, t \in I_v \\ \sum_t \alpha_t = \alpha}} \sum_{\beta_t \ge 0, t \in I_v} P_{\mbox{\scriptsize c-valid}}(\beta n) \\
\times \prod_{t \in I_v} N_{v,t}^{(\delta_t n)}(\alpha_t n, \beta_t n) \; .
\end{multline}
where $\beta = \sum_{t \in I_v} \beta_t$. Here $P_{\mbox{\scriptsize c-valid}}(\beta n)$ denotes the probability that a randomly chosen assignment of weight $\beta n$ is check-valid, and is given by
\[
P_{\mbox{\scriptsize c-valid}}(\beta n) = N_c(\beta n) \Big/ \binom{E}{\beta n} \; .
\]
Applying \cite[eqn. (25)]{di06:weight}, we find that as $n \rightarrow \infty$
\[
\binom{E}{\beta n} = \binom{n/\int \lambda}{\beta n} \rightarrow \exp  \left\{ \frac{n}{\int\lambda} h \left( \beta \int \lambda \right) \right\} \; .
\]
Combining this result with~(\ref{eq:N_c_asymptotic}), we obtain that as $n \rightarrow \infty$
\[
P_{\mbox{\scriptsize c-valid}}(\beta n) \rightarrow \exp \left\{ n Y(\beta) \right\} 
\]
where 
\[
Y(\beta) = \left( \frac{\int \rho}{\int \lambda} \right) \sum_{t \in I_c} \gamma_t \log \left( A^{(t)}(z_0) \right) - \beta \log z_0 - \frac{h(\beta \int \lambda)}{\int \lambda} \; .
\] 
Therefore, as $n \rightarrow \infty$
\begin{align}\label{eq:G_as_sum_of_exp}
\, & \mathbb{E}_{\cM_n} \left[ N_{\alpha n} \right] \rightarrow \nonumber \\
\, & \sum_{\substack{\alpha_t \ge 0, t \in I_v \\ \sum_t \alpha_t = \alpha}} 
\sum_{\beta_t, t \in I_v} \exp \left\{ n \left( \sum_{t \in I_v} X_t^{(\delta_t)}(\alpha_t, \beta_t) + Y(\beta) \right) \right\}
\end{align}
where
\begin{equation}
\beta \triangleq \sum_{t \in I_v} \beta_t \; .
\label{eq:beta_sum_of_betat}
\end{equation} 
Note that the sum in (\ref{eq:G_as_sum_of_exp}) is dominated asymptotically by the term which maximizes the argument of the exponential function. Thus, denoting the two vectors of independent variables by $\bldalpha = (\alpha_t)_{t \in I_v}$ and $\bldbeta = (\beta_t)_{t \in I_v}$, we have 
\begin{eqnarray}
G(\alpha) = \max_{\bldsmallalpha, \bldsmallbeta} S(\bldalpha, \bldbeta)
\label{eq:numerical_evaluation_1_weak}
\end{eqnarray}
where
\begin{equation}
S(\bldalpha, \bldbeta) = \sum_{t \in I_v} X_t^{(\delta_t)}(\alpha_t, \beta_t) + Y(\beta) 
\label{eq:S_definition}
\end{equation}
where $\beta$ is given by (\ref{eq:beta_sum_of_betat}), and the maximization is subject to the constraint 
\begin{equation}
R(\bldalpha, \bldbeta) = \sum_{t \in I_v} \alpha_t = \alpha 
\label{eq:sum_alpha_constraint}
\end{equation} 
together with $\alpha_t \ge 0$ and appropriate inequality constraints on $\beta_t$ for each $t \in I_v$, and $\sum_t \alpha_t = \alpha$.

Note that (\ref{eq:z0_eqn}) provides an implicit definition of $z_0$ as a function of $\bldbeta$. Similarly, for any $t \in I_v$, (\ref{eq:x0t_y0t_eqn_1}) and (\ref{eq:x0t_y0t_eqn_2}) provide implicit definitions of $x_{0,t}$ and $y_{0,t}$ as functions of the two variables $\alpha_t$ and $\beta_t$. 

We solve the constrained optimization problem using Lagrange multipliers, ignoring for the moment the inequality constraints. At the maximum, we must have
\[
\frac{\partial S(\bldalpha, \bldbeta)}{\partial \alpha_t} = \mu \frac{\partial R(\bldalpha, \bldbeta)}{\partial \alpha_t}  
\]
for all $t \in I_v$, where $\mu$ is the Lagrange multiplier. This yields
\begin{eqnarray*}
\frac{\partial x_{0,t}}{\partial \alpha_t} \left[ \delta_t \frac{ \frac{\partial B^{(t)}}{\partial x} (x_{0,t},y_{0,t})}{B^{(t)}(x_{0,t},y_{0,t})} - \frac{\alpha_t}{x_{0,t}} \right] - \log x_{0,t} \\
+ \frac{\partial y_{0,t}}{\partial \alpha_t} \left[ \delta_t \frac{ \frac{\partial B^{(t)}}{\partial y} (x_{0,t},y_{0,t})}{B^{(t)}(x_{0,t},y_{0,t})} - \frac{\beta_t}{y_{0,t}} \right] = \mu \; .
\end{eqnarray*}
The terms in square brackets are zero due to (\ref{eq:x0t_y0t_eqn_1}) and (\ref{eq:x0t_y0t_eqn_2}) respectively; therefore this simplifies to $\log x_{0,t} = -\mu$ for all $t \in I_v$.
We conclude that all of the $\{ x_{0,t} \}$ are equal, and we may write 
\begin{equation}
x_{0,t} = x_0 \quad \forall t \in I_v \; .
\label{eq:x0_all_equal}
\end{equation} 
At the maximum, we must also have
\[
\frac{\partial S(\bldalpha, \bldbeta)}{\partial \beta_t} = \mu \frac{\partial R(\bldalpha, \bldbeta)}{\partial \beta_t}  
\]
for all $t \in I_v$. This yields
\begin{multline}
\frac{\partial x_{0,t}}{\partial \alpha_t} \left[ \delta_t \frac{ \frac{\partial B^{(t)}}{\partial x} (x_{0,t},y_{0,t})}{B^{(t)}(x_{0,t},y_{0,t})} - \frac{\alpha_t}{x_{0,t}} \right] - \log y_{0,t} - \log z_0 \\
+ \frac{\partial y_{0,t}}{\partial \beta_t} \left[ \delta_t \frac{ \frac{\partial B^{(t)}}{\partial y} (x_{0,t},y_{0,t})}{B^{(t)}(x_{0,t},y_{0,t})} - \frac{\beta_t}{y_{0,t}} \right] - \log \left( \frac{1 - \beta \int \lambda}{\beta \int \lambda} \right) \\
+ \frac{\partial z_0}{\partial \beta_t} \left[ \left( \frac{\int \rho}{\int \lambda} \right) \sum_{s \in I_c} \gamma_s \frac{\frac{\mathrm{d} A^{(s)}}{\mathrm{d} z}(z_0)}{A^{(s)}(z_0)} - \frac{\beta}{z_0} \right] = 0 \; .
\end{multline}
The terms in square brackets are zero due to (\ref{eq:x0t_y0t_eqn_1}), (\ref{eq:x0t_y0t_eqn_2}) and (\ref{eq:z0_eqn}) respectively; therefore this simplifies to
\begin{equation}
z_0 y_{0,t} \left( \frac{1 - \beta \int \lambda}{\beta \int \lambda} \right) = 1 \quad \forall t \in I_v \; .
\label{eq:z0_y0_beta_1}
\end{equation}
We conclude that all of the $\{ y_{0,t} \}$ are equal, and we may write 
\begin{equation}
y_{0,t} = y_0 \quad \forall t \in I_c \; .
\label{eq:y0_all_equal}
\end{equation} 
Rearranging (\ref{eq:z0_y0_beta_1}) we obtain (\ref{eq:z0_y0_relation}). Also, summing (\ref{eq:x0t_y0t_eqn_1}) over $t \in I_v$ and using (\ref{eq:sum_alpha_constraint}) and (\ref{eq:x0_all_equal}) yields (\ref{eq:x0_y0_eqn_1}). Similarly, summing (\ref{eq:x0t_y0t_eqn_2}) over $t \in I_v$ and using (\ref{eq:beta_sum_of_betat}) and (\ref{eq:y0_all_equal}) yields (\ref{eq:x0_y0_eqn_2}). Substituting back into (\ref{eq:S_definition}) and using (\ref{eq:x0_all_equal}), (\ref{eq:y0_all_equal}), (\ref{eq:sum_alpha_constraint}) and (\ref{eq:beta_sum_of_betat}) yields 
\begin{multline}
G(\alpha) = \sum_{t \in I_v} \delta_t \log B^{(t)}(x_{0},y_{0}) - \alpha \log x_{0} - \beta \log y_{0} \\
+ \left( \frac{\int \rho}{\int \lambda} \right) \sum_{s \in I_c} \gamma_s \log A^{(s)}(z_0) - \beta \log z_0 - \frac{h(\beta \int \lambda)}{\int \lambda}
\label{eq:growth_rate_polynomial_general2}
\end{multline}
where $x_0$, $y_0$, $z_0$ and $\beta$ are the unique positive real solutions to the $4 \times 4$ system of equations (\ref{eq:z0_eqn}), (\ref{eq:x0_y0_eqn_1}), (\ref{eq:x0_y0_eqn_2}) and (\ref{eq:z0_y0_relation}). Finally, (\ref{eq:z0_y0_relation}) leads to the observation that
\[
-\beta \log z_0 - \beta \log y_0 - \frac{h(\beta \int \lambda)}{\int \lambda} = \frac{\log \left( 1 - \beta \int \lambda \right)}{\int \lambda} 
\] 
which, when substituted in (\ref{eq:growth_rate_polynomial_general2}), leads to (\ref{eq:growth_rate_polynomial_general}).

\section{Examples}
\label{sec:examples}
In this section the growth rates of two example D-GLDPC ensembles of design rate $R=1/2$ are evaluated using the polynomial solution of Theorem \ref{thm:growth_rate}. We use Hamming $(7,4)$ codes as generalized CNs and SPC codes as generalized VNs. Three representations of SPC VNs are considered, namely, the cyclic (C), the systematic (S) and the antisystematic (A) representations \footnote{The $(k \times (k+1))$ generator matrix of a SPC code in A form is obtained from the generator matrix in S form by complementing each bit in the first $k$ columns. Note that a $(k \times (k+1))$ generator matrix in A form represents a SPC code if and only if the code length $q=k+1$ is odd. For even $k+1$ we obtain a $d_{\min}=1$ code with one codeword of weight $1$.}. 

Ensemble $1$ is characterized by two CN types and two VN types. Specifically, we have $I_c = \{ 1,2 \}$, where $1 \in I_c$ denotes a $(7,4)$ Hamming CN type and $2 \in I_c$ denotes a length-$7$ single parity check (SPC) CN type, and $I_v = \{ 1,2 \}$, where $1 \in I_v$ denotes a repetition-$2$ VN type and $2 \in I_v$ denotes a length-$7$ SPC CN type in cyclic form. 
Ensemble $2$ is characterized by two CN types and four VN types. Specifically, we have $I_c = \{ 1,2 \}$, where $1 \in I_c$ denotes a $(7,4)$ Hamming CN type and $2 \in I_c$ denotes a SPC-$7$ CN type, and $I_v = \{ 1,2,3,4 \}$, where $1 \in I_v$ denotes a repetition-$2$ VN type, $2 \in I_v$ denotes a length-$7$ SPC CN type in cyclic form, $3 \in I_v$ denotes a length-$7$ SPC CN type in antisystematic form, and $4 \in I_v$ denotes a length-$7$ SPC CN type in systematic form. 
The edge-perspective type distributions of the two ensembles are summarized in Table~\ref{table:ensembles}.

\begin{figure}
\begin{center}
\psfrag{xlabel}[c]{\small{$\phantom{---}\alpha$}} \psfrag{ylabel}[c]{\small{$\phantom{---}$\textsf{Growth rate, }$G(\alpha)$}} \psfrag{Ensemble 1 (2 VN types)}[l]{\begin{picture}(0,0)\put(0.2,0.2){\scriptsize{\textsf{Ensemble $1$ ($2$ VN types)}}}\end{picture}} \psfrag{Ensemble 2 (4 VN types)}[l]{\begin{picture}(0,0)\put(0.2,0.2){\scriptsize{\textsf{Ensemble $2$ ($4$ VN types)}}}\end{picture}}
\includegraphics[%
  width=1.0\columnwidth,
  keepaspectratio]{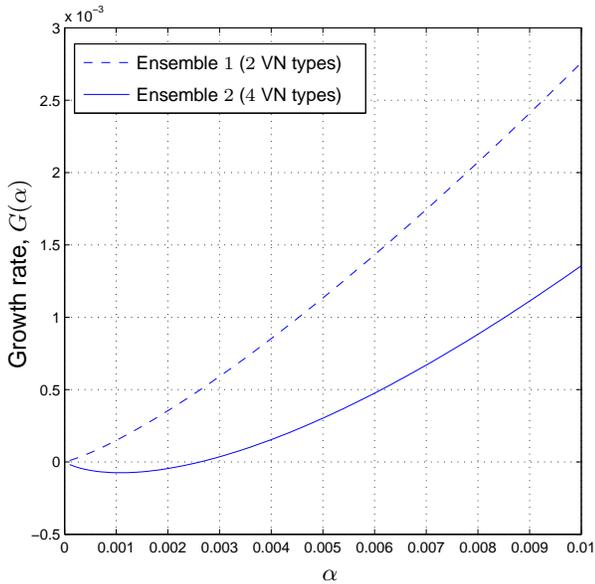}
\end{center}
\caption{\label{cap:growth_rates_paper} Growth rates of the two example ensembles described in Section \ref{sec:examples}. Ensemble $1$ is asymptotically bad, while Ensemble $2$ is asymptotically good with an ensemble relative minimum distance of $\alpha^* = 2.625 \times 10^{-3}$.}
\end{figure}

Both Ensemble 1 and Ensemble 2 have been obtained by performing a decoding threshold optimization with differential evolution (DE) \cite{storn05:differential-book}. Ensemble $1$ has been obtained by only imposing the node type and $R=1/2$ constraint. In this case we have $C \cdot V = 1.19 > 1$, so the ensemble is \emph{asymptotically bad} ($\alpha^*=0$). Ensemble $2$ has been obtained by imposing the node type and $R=1/2$ constraint, together with the constraints $C \cdot V \leq 0.5$ and $\lambda_2 \geq 0.1$. Since in this case we have $C \cdot V = 0.5 < 1$, the ensemble is \emph{asymptotically good} ($\alpha^*>0$). The expected asymptotically bad or good behavior of the two ensembles is reflected in the growth rate curves shown in Fig.~\ref{cap:growth_rates_paper}. Using a standard numerical solver, it took only $5.1$ s and $6.7$ s to evaluate $100$ points on the Ensemble $1$ curve and on the Ensemble $2$ curve, respectively. The relative minimum distance of Ensemble $2$ is \mbox{$\alpha^* = 2.625 \times 10^{-3}$}.

\begin{table}[!t]
\caption{Coefficients of $\lambda(x)$ and $\rho(x)$ for the two example D-GLDPC ensembles.}\label{table:ensembles}
\begin{center}
\begin{tabular}{llll}
\hline\hline
\multicolumn{4}{c}{\emph{Ensemble $1$}}\\
\hline
\multicolumn{2}{l}{\emph{Variable nodes}} & \multicolumn{2}{l}{\emph{Check nodes}}\\
1:repetition$-2$ & $\lambda_1=0.055646$ & 1:Hamming$(7,4)$ & $\rho_1=0.965221$\\
2:SPC$-7$ (C) & $\lambda_2=0.944354$ & 2:SPC$-7$ & $\rho_2=0.034779$\\
\hline
\multicolumn{4}{c}{\emph{Ensemble $2$}}\\
\hline
\multicolumn{2}{l}{\emph{Variable nodes}} & \multicolumn{2}{l}{\emph{Check nodes}}\\
1:repetition$-2$ & $\lambda_1=0.022647$ & 1:Hamming$(7,4)$ & $\rho_1=0.965221$\\
2:SPC$-7$ (C) & $\lambda_2=0.100000$ & 2:SPC$-7$ & $\rho_2=0.034779$\\
3:SPC$-7$ (A) & $\lambda_2=0.539920$ &  & \\
4:SPC$-7$ (S) & $\lambda_2=0.337432$ &  & \\
\hline \hline
\end{tabular}
\end{center}
\end{table}

\section{Conclusion}
A general expression for the asymptotic growth rate of the weight distribution of irregular D-GLDPC ensembles has been presented. Evaluation of the expression requires solution of a $4 \times 4$ polynomial system, irrespective of the number of VN and CN types in the ensemble. Simulation results were presented for two example optimized irregular D-GLDPC code ensembles.

\section*{Acknowledgment}
This work was supported in part by the EC under Seventh FP grant agreement ICT OPTIMIX n.INFSO-ICT-214625.

\end{document}